\def\erg{{\rm\thinspace erg}}

\def\keV{{\rm\thinspace keV}}
\def\km{{\rm\thinspace km}}

\def\Mpc{{\rm\thinspace Mpc}}
\def\Msun{\hbox{$\rm\thinspace M_{\odot}$}}

\def\s{{\rm\thinspace s}}
\def\yr{{\rm\thinspace yr}}

\def\ergps{\hbox{$\erg\s^{-1}\,$}}

\def\kmps{\hbox{$\km\s^{-1}\,$}}

\def\Msunpyr{\hbox{$\Msun\yr^{-1}\,$}}

\def\kmpspMpc{\hbox{$\kmps\Mpc^{-1}$}}
\documentstyle[psfig,times]{mn}
\begin{document}
\title{Chandra imaging of the X-ray core of Abell\,1795}
\author[]
{\parbox[]{6.in} {A.C.~Fabian, J.S.~Sanders, S.~Ettori, G.B.~Taylor$^*$,
S.W.~Allen, C.S.~Crawford, K.~Iwasawa and R.M.~Johnstone \\
\footnotesize
Institute of Astronomy, Madingley Road, Cambridge CB3 0HA \\
$*$ National Radio Astronomy Observatory, P.O. Box 0, Socorro, NM
87801, USA\\
}}

\maketitle
\begin{abstract}
We report the discovery of a 40~arcsec long X-ray filament in the core
of the cluster of galaxies A\,1795. The feature coincides with an
H$\alpha+$NII filament found by Cowie et al in the early 1980s and
resolved into at least 2 U-band filaments by McNamara et al in the
mid 1990s. The (emission-weighted) temperature of the X-ray emitting
gas along the filament is $2.5-3\keV$, as revealed by X-ray colour
ratios. The deprojected temperature will be less. A detailed
temperature map of the core of the cluster presented. The cD galaxy at
the head of the filament is probably moving through or oscillating in
the cluster core. The radiative cooling time of the X-ray emitting gas
in the filament is about $3\times 10^8\yr$ which is similar to the age
of the filament obtained from its length and velocity. This suggests
that the filament is produced by cooling of the gas from the
intracluster medium. The filament, much of which is well separated
from the body of the cD galaxy and its radio source, is potentially of
great importance in helping to understand the energy and ionization
source of the optical nebulosity common in cooling flows.

\end{abstract}


\section{Introduction}

The rich cluster Abell\,1795, at redshift $z=0.063$, has been
well-studied at optical, radio and X-ray wavelengths. It shows a
relaxed X-ray structure (Buote \& Tsai 1996) with a strong central
peak of cooler gas (Fabian et al 1994; Briel \& Henry 1996; Allen \&
Fabian 1997; Ikebe et al 1999; Allen et al 1999). This is good
evidence for a cooling flow (Fabian 1994) which is supported by the
presence of strong emission-line nebulosity around the central cD
galaxy (Cowie et al 1983; van Breugel, Heckman \& Miley 1984; Hu,
Cowie \& Wang 1985; Heckman et al 1989) accompanied by excess blue
light (Johnstone, Fabian \& Nulsen 1987; Smith et al 1997; Cardiel,
Gorgas \& Aragon-Salamanca 1998). The blue light is probably due to
massive stars (McNamara et al 1996a), with some from young globular
clusters (Holtzman et al 1996). Molecular gas in the cD has been
detected through molecular hydrogen emission (Falcke et al 1998). The
small central radio source 4C\,26.42 shows high Faraday rotation
measures (Ge \& Owen 1993) which is also a characteristic of a cooling
flow (Taylor et al 1991, 1999). The radio source may have triggered
some of the star formation in the cD (McNamara et al 1996b).

Hill et al (1988) reported that the cD in A1795 is not at rest in the
gravitational potential of the cluster but has a peculiar radial
velocity of $365\kmps$. This was supported by the outer emission-line
nebulosity having a blueshift relative to the cD galaxy. Oegerle \& Hill
(1994) later revised the velocity, and the significance of the effect,
down to about $150\kmps$, and argue that a subcluster has merged with
the cluster.

Here, we present for the first time Chandra X-ray images of the core
of A\,1795. The high spatial resolution of Chandra allows us to see an
80~kpc long filament of X-ray emission extending SSE from the central
galaxy. This filament coincides with a an H$\alpha$ emission feature
discovered by Cowie et al (1983). We discuss possible causes for such
a striking feature.

\section{The Chandra observations}

Chandra (Weisskopf et al 2000) observed A\,1795 for 19,594~s on 1999
December 20 and for 19421~s on 2000 March 21. The central galaxy was
centred 3 arcmin from the edge of the ACIS-S CCD chip, yielding good
cosmetic quality and sensitivity from about 0.3 -- 7 keV. Detailed
spectral studies of the cluster will be reported elsewhere (Ettori et
al 2000, in preparation). We report here on the X-ray structure of the
cluster core and present simple X-ray colour maps.

\begin{figure*}
\centerline{\psfig{figure=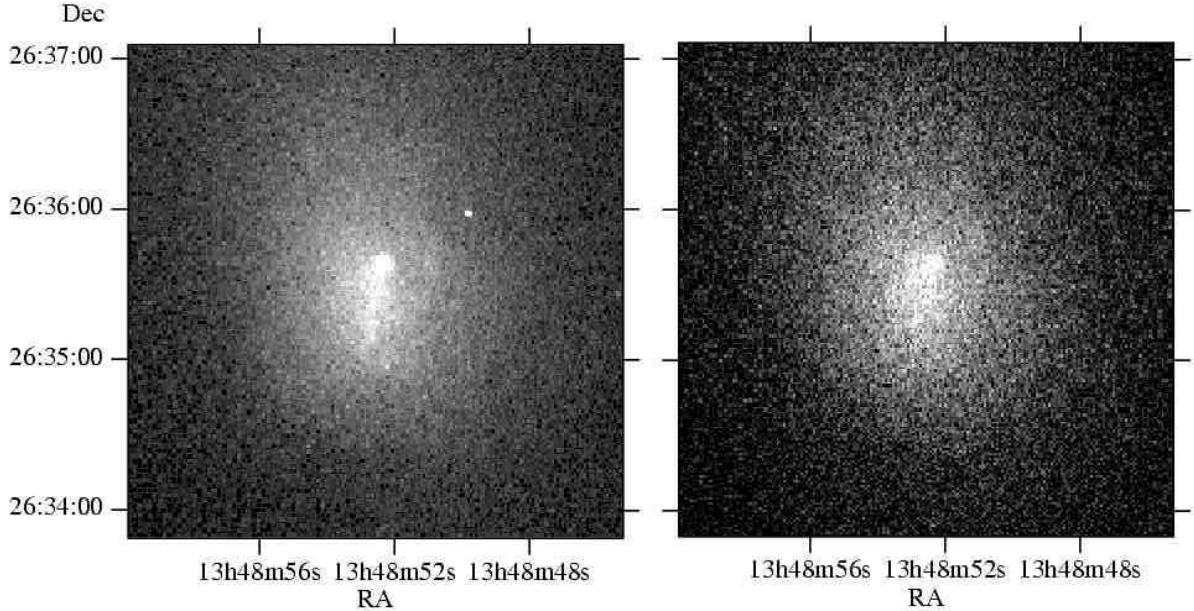,width=0.9\textwidth,angle=270}}
\caption{Soft (0.3--1.5~keV) and hard (1.5--7~keV) Chandra X-ray images of the
core of A\,1795. Pixels are one arcsec wide.}
\end{figure*}

\begin{figure}
\centerline{\psfig{figure=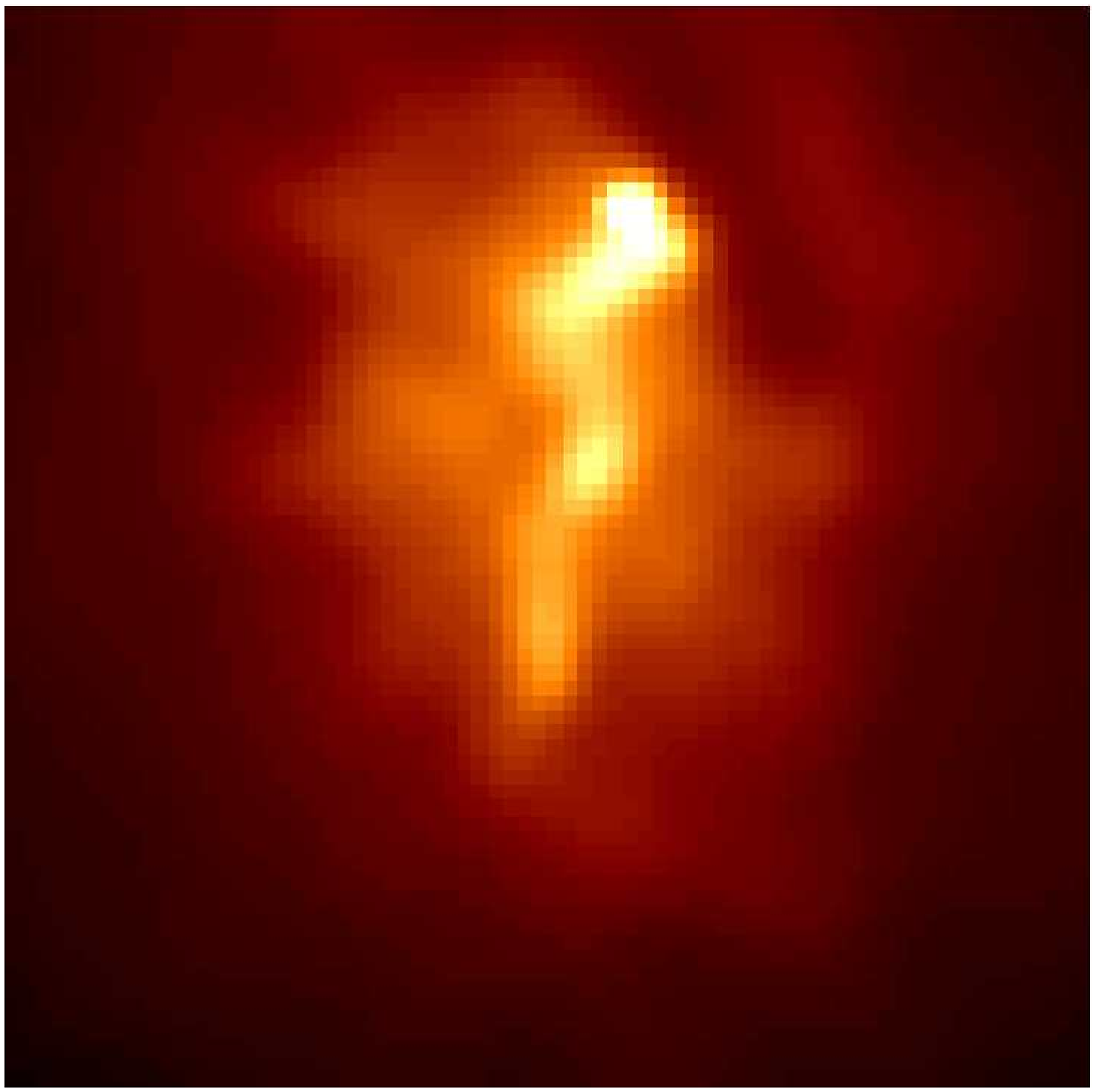,width=0.45\textwidth}}
\centerline{\psfig{figure=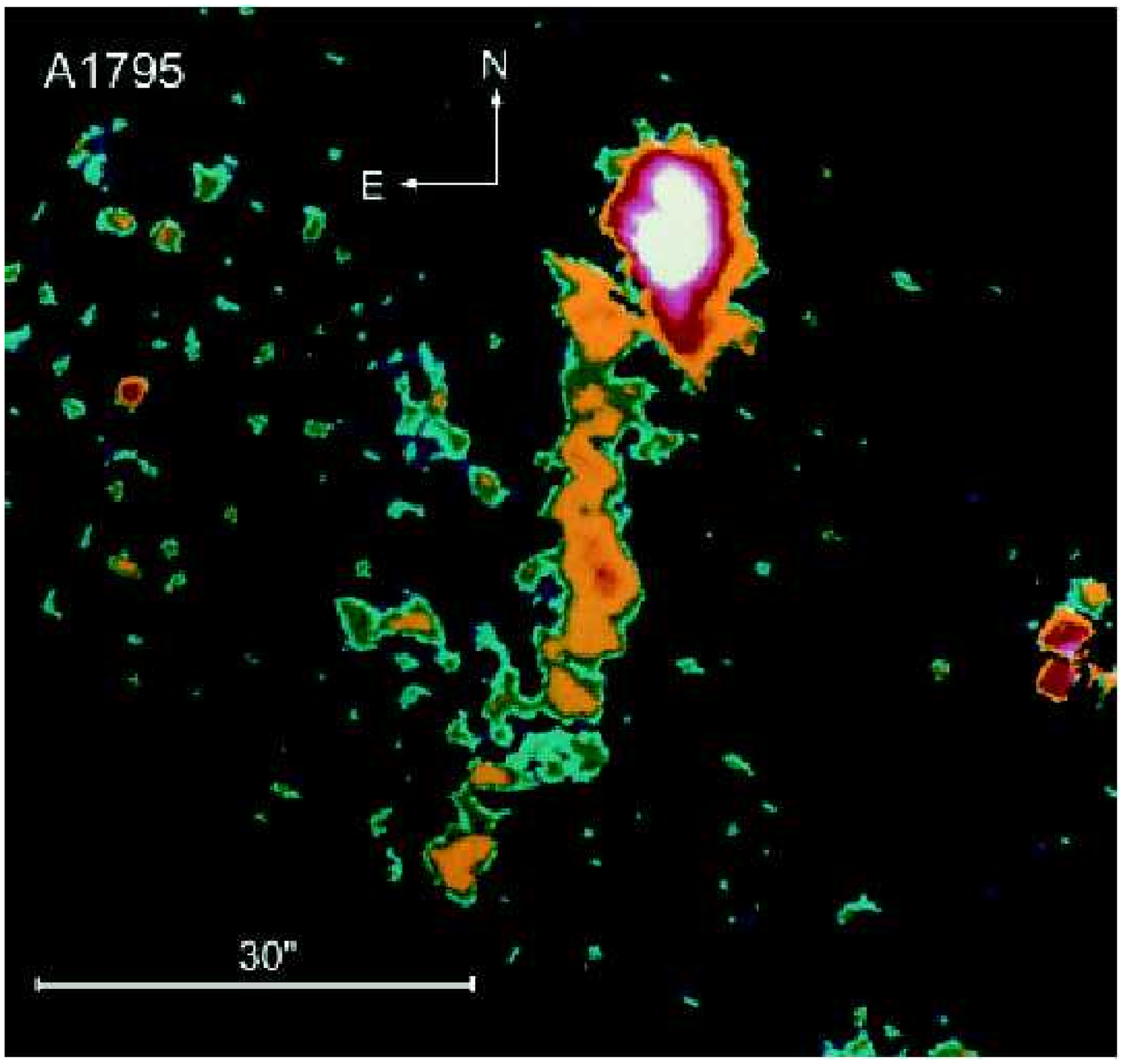,width=0.45\textwidth}}
\caption{Adaptively smoothed X-ray image (top) over the same region as the
H$\alpha+$N[II] image (below) of Cowie et al (1983). North is to the top 
and East to the left. Features seen in the X-ray map should be
significant at least at the 3$\sigma$ level.}
\end{figure}

\begin{figure}
\centerline{\psfig{figure=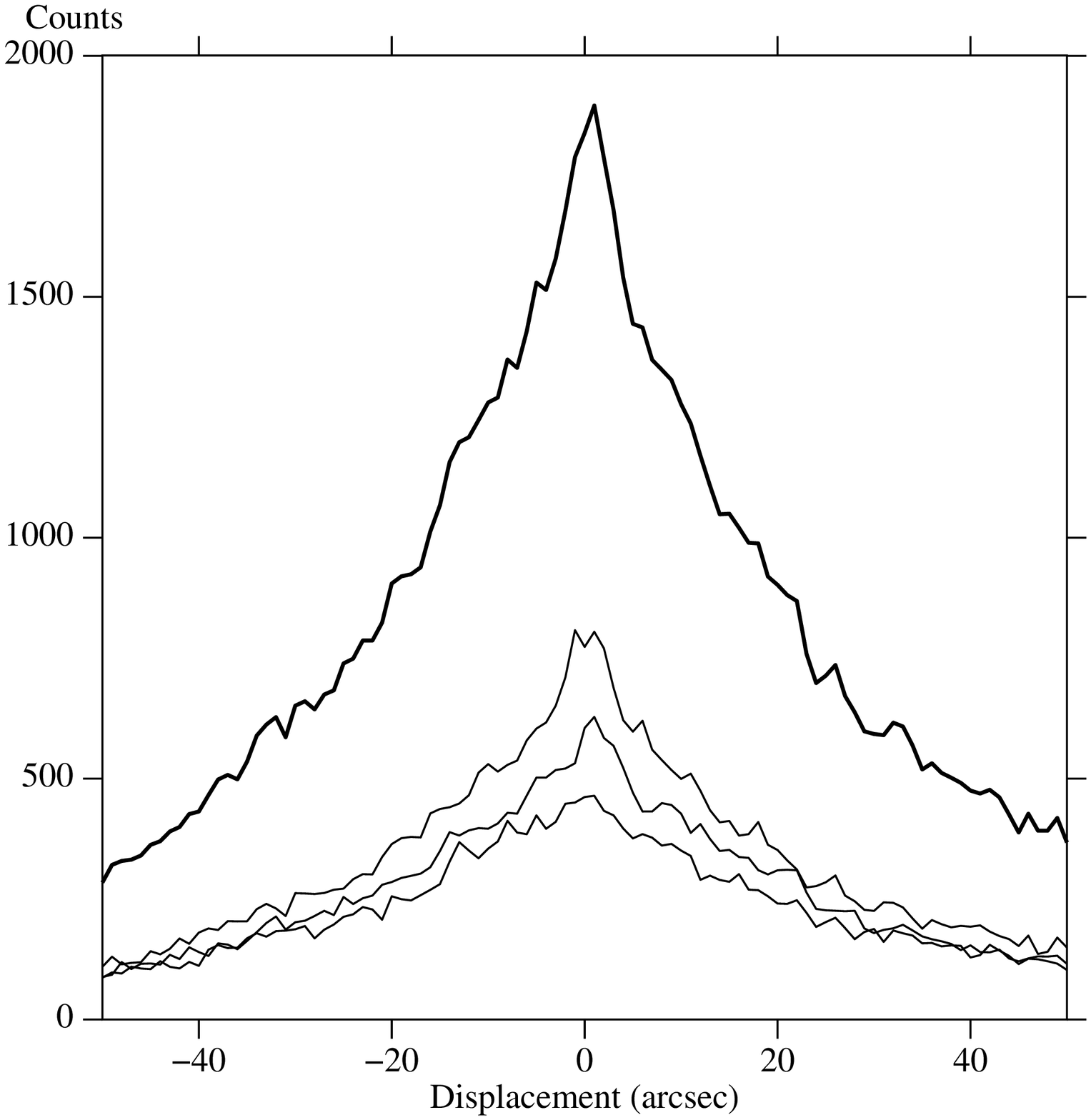,width=0.4\textwidth}}
\centerline{\psfig{figure=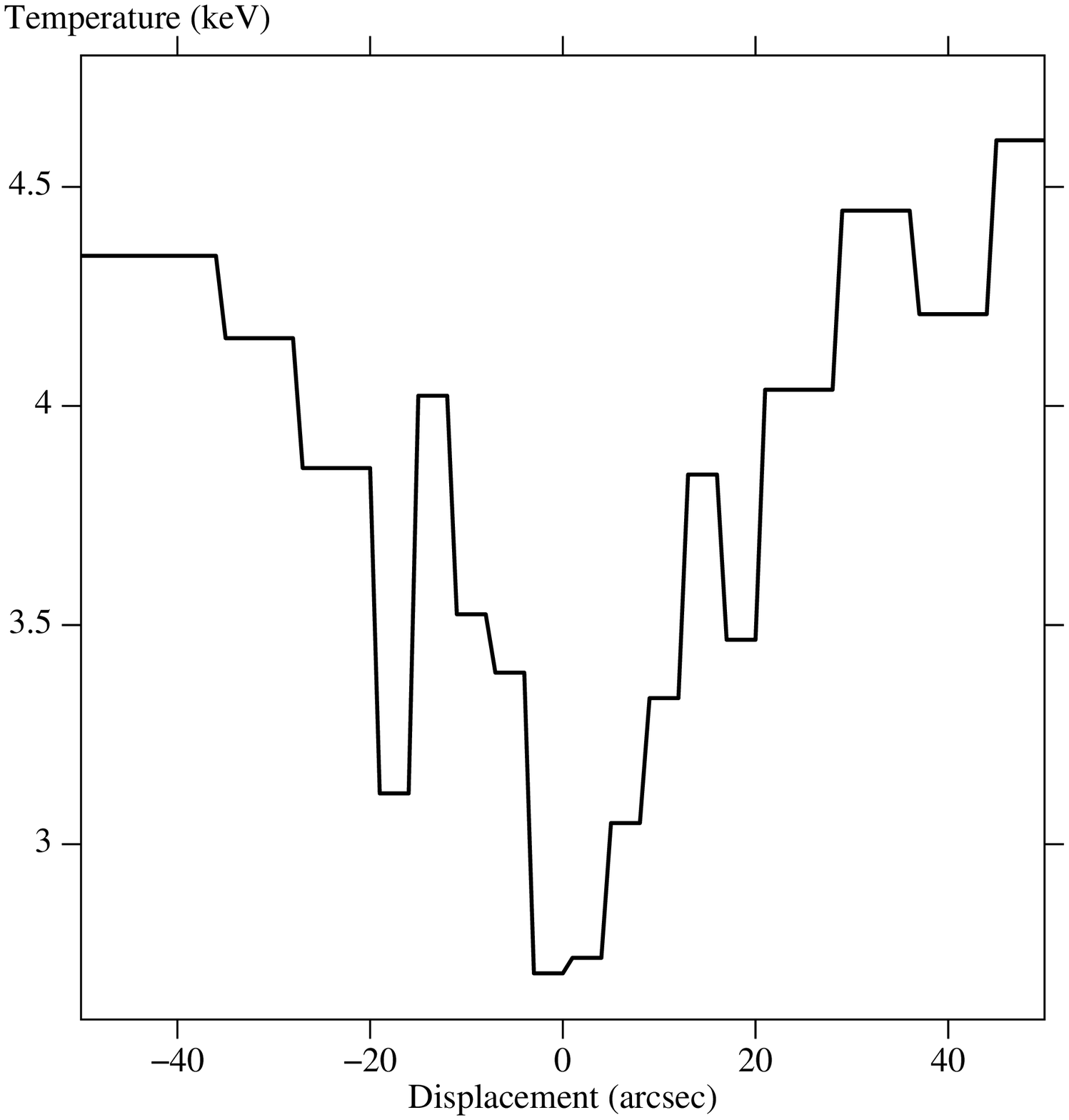,width=0.4\textwidth}}
\caption{X-ray profiles (top panel) perpendicular to the
filament in, bottom to top, the 1.5--7~keV , 0.3--0.8~keV,
0.8--1.5~keV and total bands (errors are poisson). The lower panel
shows the temperature inferred from the adaptively-binned colour
ratios (errors are about 10 per cent).}
\end{figure}

\begin{figure}
\centerline{\psfig{figure=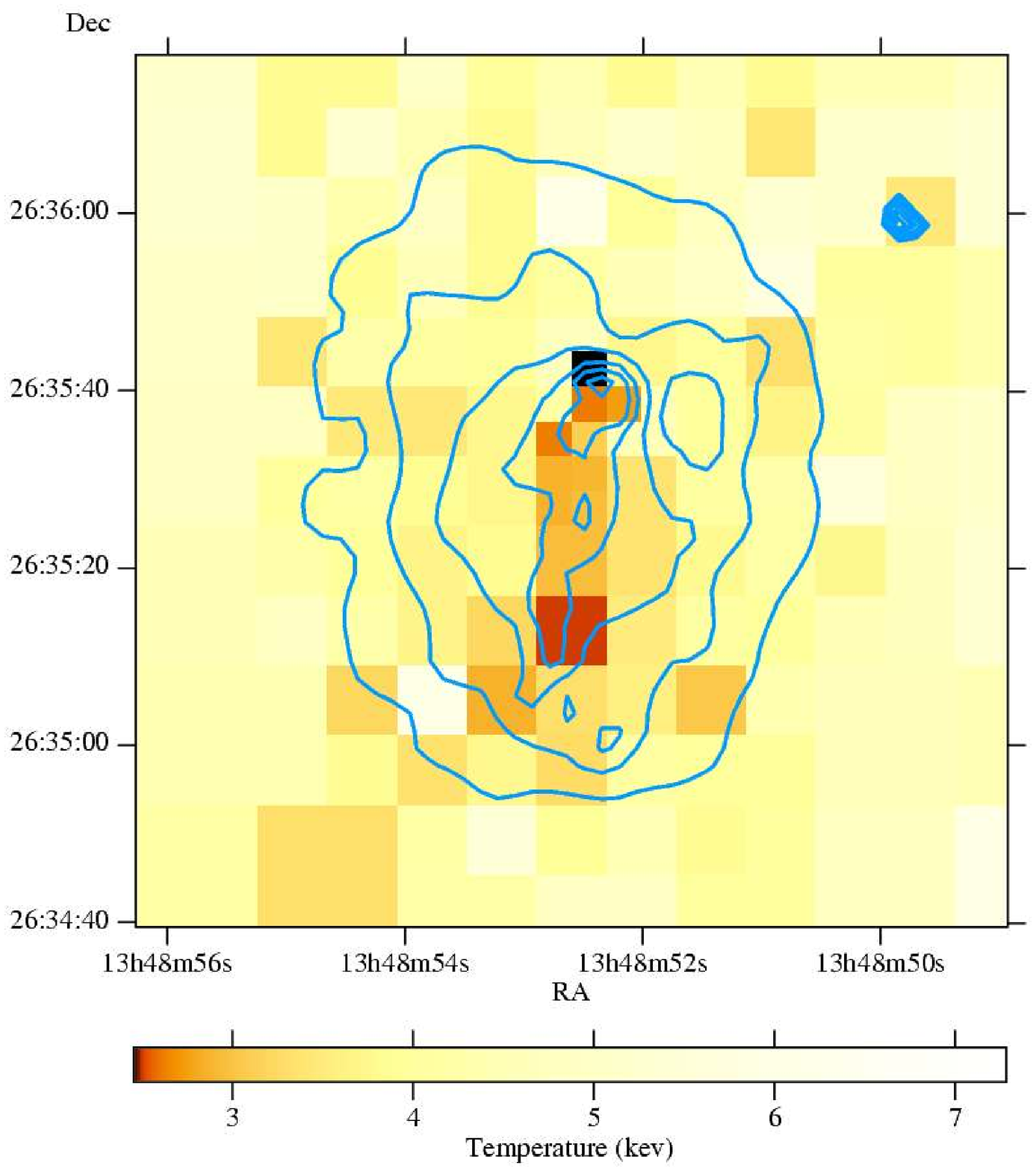,width=0.4\textwidth}}
\caption{The adaptively-binned
temperature map derived from separate colour images.}
\end{figure}

X-ray images of the cluster in the 0.3--1.5 and 1.5--7~keV bands are
shown in Fig.~1. A linear structure running N-S is apparent in the
soft image, but not in the hard one.  An adaptively smoothed (Ebeling,
White \& Rangarajan 1999) X-ray image is shown in Fig.~2 together with
a copy of the H$\alpha +$NII image of Cowie et al (1983) at
approximately the same scale. The cD galaxy lies at the top of the
filament seen in both X-rays and H$\alpha +$NII.

The X-ray profiles across the filament, in three energy bands
(0.3--0.8, 0.8--1.5 and 1.5--7~keV), are shown in Fig.~3. The bins are
$35\times 1$ arcsec, with the long axis parallel to the filament.
These have then been adaptively binned and converted to colour ratios
which have been compared with colours predicted by MEKAL thermal
models (as discussed for the Perseus cluster by Fabian et al 2000).
The bin size is varied across the image to preserve the same
signal-to-noise ratio in all bins (Sanders \& Fabian, in preparation).
This procedure has yielded the temperature profile across the filament
in Fig.~3 and the temperature map of Fig.~4. The temperature map shows
the profile along the filament and indicates that the pixel at the
Southern end of the filament is cooler ($2.5\pm 0.15\keV$) than the
remainder of the filament to the North (which is at about $2.9\pm
0.15\keV$). The coolest spot on the map coincides with the cD galaxy
position where the temperature is $2\pm 0.15\keV$. Note that these are
emission-weighted, line-of-sight, temperatures. As shown in Fig.~3,
the filament itself may only give about one third of the counts at
that position, the rest being projected into that position from the
surrounding gas. The true emission-weighted temperature of the
filament is likely to be lower than the values quoted above, probably
by about 1~keV.

Approximate electron densities have also been determined by assuming
that the pathlength through the emitting gas equals the radius. The
radiative cooling time is less than 5~Gyr within the inner arcmin
radius and about $3\times 10^8\yr$ along the filament and on the cD
galaxy. We also find evidence from the colour maps for excess X-ray
absorption at the position of the filament and cD from the colour
profile analysis.

\section{Optical and radio comparison}

The X-ray filament coincides with an H$\alpha$ filament detected
nearly two decades ago (Fig.~2).  There are no obvious galaxies which
can be responsible for this feature. McNamara et al (1996a) show
$U$-band images of this region in which the filament is well-detected
and indeed resolved into at least two separate filaments. It is at a
level which, they claim, probably requires continuum emission (e.g.
stars) in addition to line emission.

\begin{figure}
\centerline{\psfig{figure=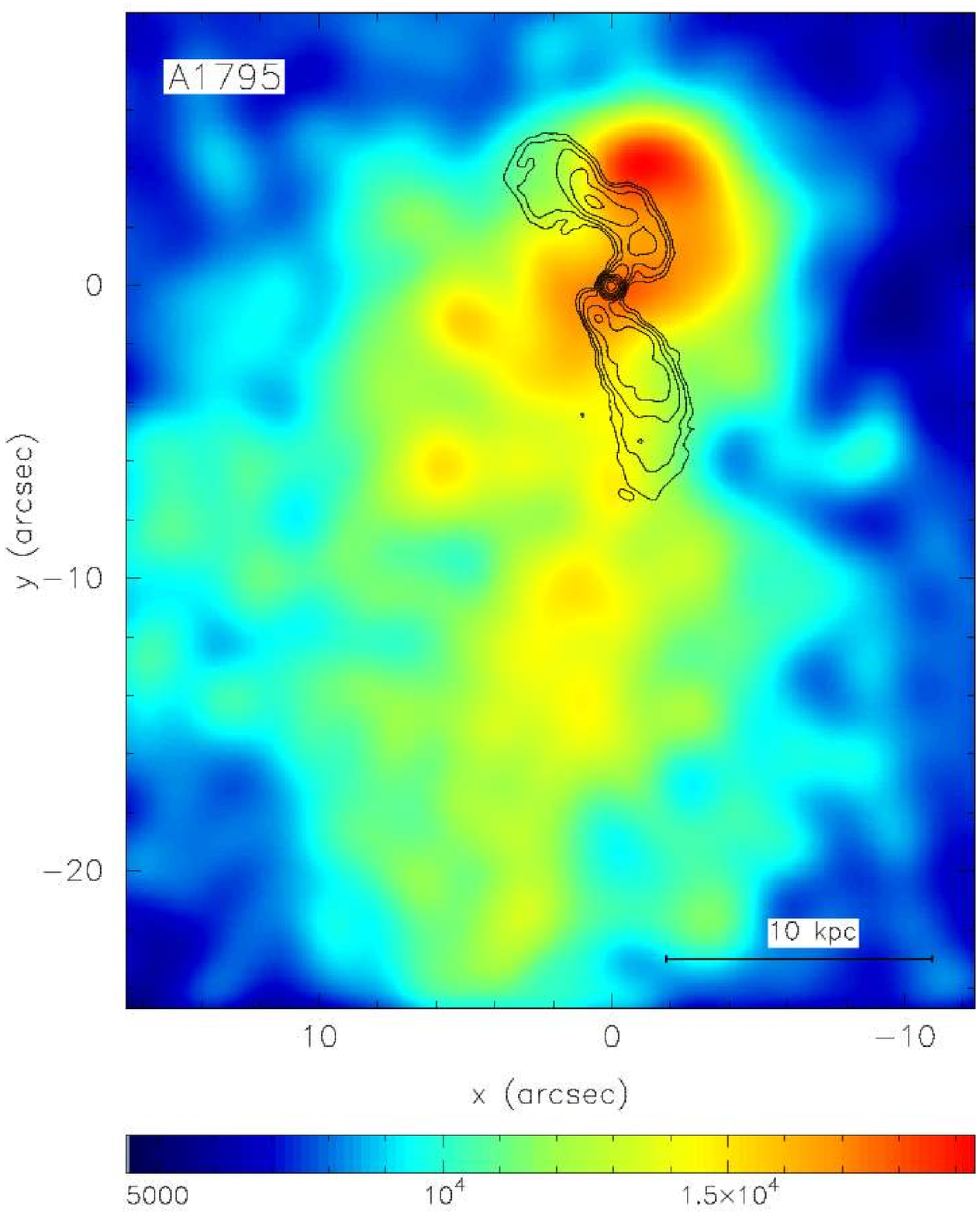,width=0.45\textwidth}}
\caption{An overlay of the 3.6 cm radio emission (Ge \& Owen 1993)
at 0.37\arcsec\ resolution on the Chandra X-ray image convolved with a
2\arcsec\ Gaussian.  The radio contours start at 0.08 mJy/beam and
increase by factors of 2 to just below the peak in the image of 33
mJy/beam. }
\end{figure}

The radio source extends about 5 arcsec to the north of the
nucleus, and has a similar extension to the south. Its orientation is
shown in Fig.~5. In Fig.~2 we can just see faint depressions in the
X-ray emission to the N and W of the present radio source. These may
perhaps be buoyant older radio lobes, as seen in the Perseus cluster
(Fabian et al 2000) The X-ray emission from the cD galaxy is seen as a
curved structure brightest at the North, about 5 arcsec from the
nucleus (Fig.~5). This morphology closely resembles that of the
H$\alpha+$NII emission there (see Fig.~6 of van Breugel et al 1984).
There is also a dust lane associated with the optically-bright regions
(Pinkney et al 1996; McNamara et al 1996b).

The polarized intensity of the radio emission is anticorrelated with
the optical line emission, presumably due to substantial Faraday
rotation within the line emitting gas (Ge \& Owen 1993). This requires
the line-emitting gas to be in front of the radio source. The
correlation between the optical line and X-ray emission in the cD
indicates that the X-ray gas also lies in front.

\section{Discussion}

The Chandra images confirm the short cooling time of the gas in the
centre of A1795 and also that the gas is cooler toward the centre. The
shortest cooling time is about $3\times 10^8\yr$ close to the cD galaxy
and along the brightest part of the filament. These results are
consistent with the presence of a strong cooling flow in the cluster
core.

We have discovered a filament of X-ray emission coincident with the
H$\alpha$ filament of Cowie et al (1983). The X-ray luminosity of the
filament, assuming a width of 5 arcsec and length of 40 arcsec
(corresponding to a projected length of about 80~kpc, assuming
$H_0=50\kmpspMpc$), is about $2\times 10^{42}$\ergps. This is only
about an order of magnitude greater than the H$\alpha+$NII luminosity
of the filament. This last luminosity probably converts to a
bolometric line luminosity comparable to (or possibly greater than)
the X-ray luminosity when all optical and UV emission lines are taken
into account. If there is no heating balancing cooling then the mass
cooling rate in the filament alone is about $10\Msunpyr$. The X-ray
brightest region in the cD galaxy also coincides with the
H$\alpha+$NII brightest region there.

The age of the filament given its projected length and radial velocity
of $\sim 150$\kmps corresponds to $5\times 10^8$\yr, if it is at an
angle of 45 deg to the line of sight. The timescale is of course less
if the filament lies closer to the plane of the Sky. This raises the
possibility that its length is constrained by cooling. We have found
(Fig.~4) that the coolest part of the X-ray filament is near the end
opposite to the cD galaxy. The straightness of the filament argues
against the intracluster medium being very turbulent.

We note that the radial velocity of the cD relative to the velocity
dispersion of the cluster ($\sigma = 920\kmps$, Oegerle \& Hill 1994)
is small. This suggests that if the angle of its motion to the line of
sight is low then the cD must be very close to the turnaround point in
its orbit. An approximate estimate for the difference between the
position of the cD, $\Delta r$, from the turnaround radius of its
(radial) orbit, $r$,
assuming that the core is of constant density out to radius $a,$ is
given by $\Delta r/r\approx 0.33 (v/\sigma)^2(a/r)^2.$ If the angle is
large (close to the plane of the Sky) then the restriction is much
less and the true velocity much higher.

Although the filament is clear in both the X-ray and H$\alpha+$NII
images the brightness variations along it do not compare well. The
brightest part of the H$\alpha+$NII filament about 25 arcsec S of the
centre of the cD galaxy corresponds to a minimum in the X-ray surface
brightness. The reverse is true of the X-ray bright part almost 20
arcsec from the cD. There is also little evidence in the X-ray image
for the SE pointing tail to the filaments seen in H$\alpha+$NII and
the $U$-band.

We note that the centroid of the outer X-ray emission (Fig.~4) lies
not on the cD but partway (10--20 arcsec S of the cD) along the
filament. This suggests that the cD is oscillating about the cluster
core. There is no strong evidence from the X-ray surface brightness
for a deep cusp at the cluster centre as determined from the outer
emission. This argues against any static dark matter cusp at the
cluster centre. The lack of any extensive ($>10$~arcsec) peak around
the cD also argues against any deep dark matter well associated with
the massive cD halo seen in the cluster (Johnstone, Naylor \& Fabian
1991) from moving directly with the cD itself.

There are several possible explanations for the filament. Detailed
modelling is beyond the scope of this Letter. It may a) be a cooling
wake, produced by a cooling flow occurring around a moving galaxy
(e.g. NGC5044; David et al 1994), b) a contrail produced by the ram
pressure of the radio source passing through a multiphase medium, c)
evaporation of cold gas ram-pressure stripped from the cD, or d) an
accretion wake (Sakelliou, Merrifield \& McHardy 1996).

This last possibility, d), requires that the gas be focussed by the
gravitational effect of the cD and then cool into a wake. Since the
likely space velocity and the velocity dispersion of the cD are both
at least a factor of two smaller than the sound speed of the hot gas
it is unlikely that gravitational focussing of the hot gas alone is
responisble. The evaporation scenario c) requires that there was more
than $5\times 10^9\Msun$ of gas (the hot gas mass of the filament)
within the cD to be stripped out. It is not clear that it would be
pulled out into such a straight, even filament (the
ram-pressure-stripped wake behind the elliptical galaxy M86 in the
Virgo cluster, Forman et al 1979, is very broad). Note that it is most
unlikely that there is, or has been, any significant ($\gg10^9\Msun$)
hot interstellar medium intrinsic to the cD galaxy. The radiative
cooling time of any 1~keV component in the galaxy would be only
$10^8\yr$ or less and it is replenished from stellar mass loss at only
a few $\Msunpyr$ (Takeda, Nulsen \& Fabian 1984). The cool gas at the
position of the cD is then due to a combination of stellar mass loss
and inflow from the intracluster medium.

We cannot rule out the contrail, b), but it is likely that the passage
of the galaxy would heat gas rather than lead to it cooling. The
simplest explanation is a cooling wake, a). The basic picture then is
that the hot gas is in almost hydrostatic equilibrium in the cluster
core potential, and cooling radiatively very slowly and steadily. The
gas with a cooling time exceeding $10^9\yr$ only senses the mean
gravitational field of the cD, which then averages to a line along its
direction of motion. Gas with shorter cooling time cools onto that
line as the cD passes. Basically the mass of the cD serves to focus
the inner part of the cooling flow, which is probably multiphase.
Cooling will limit the length of the wake, in accord with the coolest
X-ray gas being observed to be most distant from the cD. Of course, if
the cD is also oscillating about the cluster dark matter centre then 
the filament length will not exceed the oscillation length.

The emission-line velocity map presented by Hu et al (1985) shows that
most of the filament has the velocity of the cluster (i.e. $\sim
150\kmps$ blueshifted from the velocity of the cD; the cD is
redshifted relative to the cluster by the same amount, Oegerle \& Hill
1994). Only the line emission at the position of the cD shares its
redshift. This velocity structure agrees with the above cooling model
and argues strongly against the stripping hypothesis, where a strong
velocity gradient is expected.

The extensive X-ray/H$\alpha$ filament in A1795 offers the possibility
to identify the source of energy and ionization of the optical
nebulosity common in cooling flow clusters (Crawford et al 1999 and
references therein; Donahue et al 2000). The filament extends at least
80 kpc from the cD and most of it is therefore well removed from the
cD nucleus, the radio source, and the bulk of the stars in the cD. If
part of the $U$-band emission (McNamara et al 1996a) is due to massive
young stars then the formation of intergalactic stars can be studied.
It will clearly be interesting to observe the clumpiness of the
emission. As reported above, the optical emission from the filament
does not appear to correlate in detail with the X-ray emission, which
could be due to clumpiness in the gas: the denser parts cool out
fastest and so form stars more readily. 

The Balmer decrement of 3--4 reported by Hu et al (1985) in the
filament is consistent with a small amount of reddening. Further
optical spectra can search for emission lines from calcium (Donahue \&
Voit 1993; Fabian, Johnstone \& Daines 1994) which depletes onto dust.
If dust is present, it need not rule out condensation from the
intracluster medium since large grains may survive sputtering and
supernovae from massive stars in the filament can inject dust. This
could be helpful in studying prompt dust formation from young stellar
populations which is not clearly understood at present (Feigelson,
private communication).

The filament also enables studies of the survivability of cool gas in
the intracluster medium against mixing and conduction. Mixing layers
(Begelman \& Fabian 1990) may provide some of the photoionizing
radiation for the cold gas in the filament. Thermal conduction, if it
proceeds at the Spitzer value, would destroy the filament in $\sim
10^8\yr$, so it must be suppressed by at least an order of magnitude,
presumably by magnetic fields.

In summary, the optically-derived velocity structure of the filament
associates it with the intracluster gas, not the galaxy. The
X-ray-derived temperature structure of the filament, with the gas
furthest from the galaxy (therefore oldest) being the coolest,
indicates that the gas is cooling with time and not heating up
(evaporating). Together with the approximate match in timescales, the
evidence points to the filament originating from cooling of the
intracluster gas, attracted into a wake along the path of the moving
cD galaxy.

\section{Acknowledgements}
ACF is grateful to NASA for the opportunity to participate as an
InterDisciplinary Scientist and the Chandra project for such a superb
instrument. ACF, CSC and SWA thank the Royal Society for support.


\begin{thebibliography}{}
\bibitem []{} Allen S.W., Fabian A.C., 1997, MNRAS, 286, 583
\bibitem []{}  Allen S.W., Fabian A.C., Johnstone R.M., Nulsen P.E.J.,
Arnaud K.A., 1999, MNRAS, submitted, astro-ph/9910188
\bibitem []{} Begelman M.C., Fabian A.C., 1990, MNRAS, 244, 26P
\bibitem []{} Briel U.G., Henry J.P., 1996, ApJ, 472, 131
\bibitem []{} Buote D., Tsai J., 1996, ApJ, 458, 27
\bibitem []{} Cardiel N., Gorgas J., Aragon-Salamanca A., 1998, MNRAS,
298, 977
\bibitem []{} Cowie L.L., Hi E.M., Jenkins E.B., York D.G., 1983 ApJ,
272, 29
\bibitem []{} Crawford C.S., Allen S.W., Ebeling H., Edge A.C., Fabian
A.C., 1999, MNRAS, 306, 875 
\bibitem []{} David L.P., Jones C., Forman W., Daines S.J., 1994, ApJ,
428, 544 
\bibitem []{}  Donahue M., Voit G.M., 1993, ApJ, 414, L17
\bibitem []{}  Donahue M., Mack J., Voit G.M., Sparks W., Elston R.,
Maloney P.R., ApJ, 2000, in press (astro-ph/0007062)
\bibitem []{} Ebeling H., White D.A., Rangarajan V., 1999, MNRAS, in
press
\bibitem []{}  Fabian A.C., Hu E.M., Cowie L.L., Grindlay J., 1981,
ApJ, 248, 47
\bibitem []{}  Fabian A.C., 1994, ARAA, 32, 277
\bibitem []{} Fabian A.C., Arnaud K.A., Bautz M.W., Tawara Y., 1994,
ApJ, 436, L63
\bibitem []{} Fabian A.C., Johnstone R.M., Daines S.J., 1994b, MNRAS,
271, 737
\bibitem []{} Fabian A.C., et al 2000, MNRAS, in press
(astro-ph/0007456)
\bibitem []{} Falcke H., Rieke M.J., Rieke G.H., Simpson C., Wilson
A.S., 1998, ApJ, 494, L155
\bibitem []{} Forman W., Schwarz J., Jones C., Liller W., Fabian A.C.,
1979, ApJ, 234, L27
\bibitem []{} Ge J.P., Owen F.N., 1993, AJ, 105, 778
\bibitem []{}  Heckman T.M., Baum S.A., van Breugel W.J.M., McCarthy
P., 1989, ApJ, 338, 48
\bibitem []{} Hill J.M., Hintzen P., Oegerle W.R., Romanishin W.,
Lesser M.P., Eisenhamer J.D., Batuski D.J., 1988, ApJ, 332, L23
\bibitem []{} Holtzman J.A., et al 1996, AJ, 112, 416
\bibitem []{} Hu E.M., Cowie L.L., Wang Z., 1985, ApJS, 59, 447 
\bibitem []{} Johnstone R.M., Fabian A.C., Nulsen P.E.J., 1987, MNRAS,
224, 75 
\bibitem []{} Johnstone R.M., Naylor T., Fabian A.C., 1991, MNRAS,
248, 18P
\bibitem []{}  McNamara B.R. et al 2000, ApJ submitted, astro-ph/0001402
\bibitem []{} McNamara B.R. Januzzi B.T., Elston R., Sarazin C.L.,
Wise M., 1996, ApJ, 469, 66
\bibitem []{} McNamara B.R., Wise M., Sarazin C.L., Januzzi B.T.,
Elston R., 1996, ApJ, L9
\bibitem []{} Oegerle, W.R., Hill, J.M., 1994, AJ, 107, 857
\bibitem []{} Pinkney J. et al 1996, ApJ, 468, L13
\bibitem []{} Sakelliou I., Merrifield M.R., McHardy I.M., 1996,
MNRAS, 283, 673
\bibitem []{} Smith E.P. et al 1997, ApJ, 478, 516
\bibitem []{} Takeda H., Nulsen P.E.J., Fabian A.C., 1984, MNRAS, 208,
261
\bibitem []{} Taylor G., Barton E.L., Ge, J.-P., 1994, AJ, 107, 1942
\bibitem []{} Taylor G.B., Fabian A.C., Allen S.W., 1999, in 'Diffuse
thermal and relativistic plasma in galaxy clusters', eds H.Bohringer,
L. Feretti, P. Schuecker. MPIfE, Garching, Germany, p77
\bibitem []{} van Breugel W., Heckman T., Miley G., 1984, ApJ, 276, 79
\bibitem []{} Weisskopf M.C., Tananbaum H.D., Van Spebroeck L.P.,
O'Dell S.L., 2000, Proc SPIE, 4012, in press astro-ph/0004127
\bibitem []{} Xu, H., Makishima K., Fukazawa Y., Ikebe Y., Kikuchi K.,
Ohashi T., Tamura T., 1998, ApJ, 5000, 738

\end{thebibliography}
\end{document}